\documentstyle[12pt,epsfig]{article}
\def\be{\begin{equation}} \def\ee{\end{equation}} \def\bea{\begin{eqnarray}}
\def\eea{\end{eqnarray}} \def\nnb{\nonumber}

\begin{document}

\hfill{July 17, 2012,\ \ \ {\tt v2346}}


\begin{center}
\vskip 5mm 
\vskip 5mm 
{\Large\bf
Effective range corrections from  
effective field theory 
with di-baryon fields and perturbative pions
}
\vskip 5mm 
\vskip 5mm 
{\large Shung-Ichi Ando\footnote{sando@daegu.ac.kr}
and Chang Ho Hyun\footnote{hch@daegu.ac.kr}
}

\vskip 5mm

{\large \it 
Department of Physics Education,
Daegu University, Gyeongsan 712-714,
Republic of Korea
}

\vskip 5mm
\vskip 5mm

\end{center}

\vskip 2mm

Contributions of perturbative pions around a non-trivial fixed point 
are studied by utilizing di-baryon fields. 
We calculate ${}^1S_0$ and ${}^3S_1$ phase shifts for the $np$ 
scattering at low energies up to one-pion-exchange contributions. 
We also calculate effective range parameters, $v_2$, $v_3$, and $v_4$, 
in the higher order of the effective range expansion,
and obtain corrections of the effective range 
to the expressions previously reported by Cohen and Hansen. 
After the scattering length and the effective range are renormalized,
we study the role of renormalization scale parameter $\mu$
and we discuss the range of validity of the theory.  

\vskip 5mm \noindent
PACS(s): 11.10.Gh, 13.85.Dz, 21.45.Bc 

\newpage \noindent 
{\bf 1. Introduction} 
\vskip 1mm

The renormalization group analysis of two-nucleon interaction 
by Birse, McGovern, and Richardson~\cite{bmr-plb99} 
reveals that there are two fixed points, 
where no scales appear 
and a perturbative expansion is possible 
in the scale free limit.
Around one fixed point, so called the trivial fixed point 
at which all the interactions vanish,
a usual perturbation theory such as QED, 
chiral perturbation theory (ChPT) for meson sector 
and one nucleon sector~\cite{s-anp03,b-ppnp08}
can be constructed.
Around the other fixed point, so called the non-trivial fixed point,
the inverse of a scattering length or a binding energy 
vanishes for two-nucleon systems,
and an effective range expansion (ERE)~\cite{b-pr49,bl-pr50}
has been known for a long time in this limit.

After S. Weinberg had suggested an application of 
ChPT to nuclear physics, 
especially to nuclear forces two decades ago~\cite{w-plb90,w-npb91}, 
a lot of works have been done so far.
(For reviews, see, e.g., 
Refs.~\cite{bk-arnps02,ehm-rmp09,me-pr11} 
and references therein.\footnote{
There have been intensive discussions concerning 
renormalization and counting rules, with a finite or infinite cutoff, 
in solving the Lippmann-Schwinger equation with nucleon-nucleon interactions
obtained in chiral effective theory.
For more details, see a recent review, {\it e.g.},
Ref.~\cite{me-pr11}, and references therein. 
}
)
Because pions appear as (massless) Goldstone bosons due to 
spontaneous breaking of chiral symmetry of QCD
(finite pion mass comes out due to explicit chiral symmetry breaking 
terms, i.e., quark masses), the pions are expected to play a dominant 
role in the long-range interactions.
However, a difficulty of systematic application 
of ChPT to the two-nucleon system  
stems from the appearance of small scales such
as large $S$-wave scattering length and small binding 
momentum 
of the deuteron,
which are much smaller than the pion mass, $m_\pi\simeq 140$ MeV. 
A fine-tuning is required in reproducing those small scales 
in the $S$-wave two-nucleon systems,
which implies that the low energy constants
can be important as well~\cite{k-npa99,bbsk-npa02}.

In this work, 
we revisit the $S$-wave nucleon-nucleon scattering 
with perturbative pions proposed by Kaplan, Savage, and Wise 
(KSW)~\cite{ksw-plb98,ksw-npb98}.
We adopt the limit of the non-trivial fixed point 
as the ``fine-tuned" 
ground
state of the two nucleon system, 
utilize di-baryon fields~\cite{k-npb97,bs-npa01,ah-prc05,st-prc08}, 
which have the same quantum numbers
of two-nucleon $^1S_0$ and $^3S_1$ states
and the infinite scattering length or the zero binding energy, 
as a zeroth order two-nucleon core,
and incorporate pion clouds perturbatively
(one-pion-exchange contribution in this work 
as the first chiral correction\footnote{See footnote \ref{footnote;pions}.})
around the non-trivial fixed point being represented 
by the di-baryon fields.

This may not be so unrealistic choice because 
it was conjectured by Braaten and Hammer that
``Small increases in the up and down quark masses of QCD 
would tune the theory to the critical renormalization group
trajectory for an infrared limit cycle in the three-nucleon
system. At critical values of the quark masses, the deuteron
binding energy goes to zero and the triton has infinitely many
excited states with an accumulation point at the three-nucleon
threshold''\cite{bh-prl03}. 
Subsequently, a study with an updated chiral effective theory reported
an estimated critical pion mass, which makes 
binding energies of the deuteron and its isospin partner zero,
tuned to be
$m_\pi^{crit}\simeq 198\,$MeV \cite{ehmn-epjc06}. 
Thus the physical pion mass is closer to the critical pion mass 
than 
the mass in the chiral limit.

In addition, as shown in Ref.~\cite{bb-prc03},
because the expansion around the non-trivial 
fixed point is equivalent to ERE and 
takes a form of series of the inverse of potential 
(or the on-shell K matrix), $1/V$, 
when one renormalizes coefficients of 
the ordinary contact interactions 
by using the effective range parameters,
one has complicated nonlinear relations of 
the coefficients. 
On the other hand, 
because our expansion scheme matches well to ERE
we obtain a linear relation among the coefficients
if we include the di-baryon fields in the formalism
up to the order we considered in this work.
Thus the formalism with di-baryon fields provides us a way
simple and transparent to account for the role of one-pion
exchange perturbatively in studying the chiral corrections
in nuclear systems at low energies.

We calculate the $S$-wave phase shifts of the $NN$ 
scattering for $^1S_0$ and $^3S_1$ channels up to 
one-pion exchange contributions in the 
 framework
mentioned above,\footnote{
We basically employ the standard counting rules suggested by 
KSW~\cite{ksw-plb98,ksw-npb98}. 
However, as to be mentioned, we do not calculate all of the contributions
up to next-to-leading order (NLO) in the KSW counting rules,
and expand the one-pion-exchange corrections around the inverse of the
LO amplitude, $A_d^{-1}$.
Thus two-pion-exchange contributions which would be important are 
counted as next-to-next-to leading order (NNLO) corrections.
}
where $D$-wave mixture in the spin triplet channel is neglected
for simplicity.\footnote{
\label{footnote;pions}
It is known that the one-pion exchange potential has 
a non-perturbative short range contribution, 
the strong attractive tensor force, in $^3S_1$-$^3D_1$ 
channel~\cite{bbsk-npa02},
and the failure of the perturbative treatment of the one-pion 
exchange contribution in the higher order is reported 
in Ref.~\cite{fms-npa00}.
However, this difficulty can be avoided 
by introducing a regulator to the singular part of 
the one-pion exchange potential~\cite{bkv-prc09}.   
}
We also calculate higher order terms of the effective range corrections,
$v_2$, $v_3$, and $v_4$, whose definitions will be given below.
Those terms have been calculated by 
Cohen and Hansen~\cite{ch1-prc99,ch2-prc99} with a pionful 
effective field theory (EFT).
In this work, we obtain new corrections to them from the effective ranges. 
After the scattering length and the effective range 
are renormalized by using empirical values, 
we
retain the $\mu$-dependence in the effective range corrections,
$v_2$, $v_3$, and $v_4$, and 
study the role
of the scale parameter $\mu$ in the dimensional regularization (DR)
and power divergence subtraction (PDS) scheme.

Usually, in a perturbative calculation,
the same couplings appearing in higher orders 
as those being renormalized in lower orders by physical quantities,
the scattering length and the effective range in the present case, 
are replaced by the renormalized couplings,
and thus explicit $\mu$-dependence can be removed from 
an amplitude. 
However, the higher order contributions in the $S$-wave
nucleon-nucleon scattering amplitude converge slowly and 
would bring a significant modification to the amplitude,
thus, in this work we rather retain the $\mu$-dependence
in the higher orders.
In other words, the renormalized amplitude is 
perturbatively scale-independent, 
and adjusting the scale parameter $\mu$ 
can be regarded as a probe of ``optimal higher order contributions''.
In general, terms corresponding to 
the scattering length, the effective range, 
and the higher order corrections, 
$v_2$, $v_3$, and $v_4$, in the calculation 
are functions of the scale parameter $\mu$.
After renormalization of the scattering length 
and the effective range, we find that  
the higher order corrections, $v_2$, $v_3$ and $v_4$, are 
significantly sensitive to the value of $\mu$, and the sensitivity
has a strong correlation to the range 
where the theory has predictive power.

This work is organized as the following.
In section 2, we review pionless effective theory with 
di-baryon fields and briefly discuss relations between 
the formalism and the non-trivial fixed point. 
In section 3, we calculate the nucleon-nucleon scattering amplitudes 
for $^1S_0$ and $^3S_1$ channel
with the one-pion-exchange contributions.
In section 4, the amplitudes are renormalized by
using the empirical values of the 
scattering lengths and the effective ranges.
In section 5, the higher order effective range corrections,
$v_2$, $v_3$, and $v_4$, are obtained, and the $\mu$ 
dependences of the effective range parameters and 
the phase shifts are studied.
Finally, in section 6, discussion and conclusions are 
presented.

\vskip 3mm \noindent 
{\bf 2. Pionless theory with di-baryon fields}

In this section, we briefly review a pionless theory
with di-baryon fields. 
A relevant effective Lagrangian reads~\cite{bs-npa01,ah-prc05}
\bea
{\cal L} &=& {\cal L}_N 
+ {\cal L}_s
+ {\cal L}_t\,,
\eea
where ${\cal L}_N$ is the Lagrangian for standard one nucleon sector,
and ${\cal L}_s$ 
and ${\cal L}_t$ are those 
for the di-baryon fields 
in the $^1S_0$ and the $^3S_1$ states, respectively.
The Lagrangian ${\cal L}_N$ in the heavy-baryon formalism reads
\bea
{\cal L}_N&=& N^\dagger \left[
iv\cdot \partial 
+ \frac{(v\cdot \partial)^2 -\partial^2}{2 m_N}
\right] N\,,
\eea
where $v^\mu$ is the velocity vector satisfying the condition 
$v^2=1$, and $m_N$ is the nucleon mass. 
The Lagrangians with the di-baryon fields read   
\bea
{\cal L}_s &=& \sigma_s s_a^\dagger \left[ 
iv\cdot \partial 
+ \frac{(v\cdot \partial)^2 -\partial^2}{4 m_N}
+ \Delta_s \right] s_a
- y_s \left[s_a^\dagger (N^TP^{(^1S_0)}_aN) 
+ \mbox{\rm h.c.}
\right] ,
\label{eq;Ls}
\\
{\cal L}_t &=& \sigma_t t_i^\dagger \left[ 
iv\cdot \partial 
+ \frac{(v\cdot \partial)^2 -\partial^2}{4 m_N}
+ \Delta_t \right] t_i
- y_t \left[t_i^\dagger (N^TP^{(^3S_1)}_iN) 
+ \mbox{\rm h.c.}
\right] ,
\label{eq;Lt}
\eea
where $s_a$ and $t_i$ are the di-baryon fields for the 
$^1 S_0$ and $^3 S_1$ states, respectively, $\sigma_{s,t} = -1$,
$\Delta_{s,t}$ are mass differences between the di-baryon masses and 
the two nucleon mass, and $y_{s,t}$ are coupling constants of 
di-baryon-nucleon-nucleon ($dNN$) vertices. 
$P^{(^1S_0)}_a$ and $P^{(^3S_1)}_i$ are projection operators 
for $^1S_0$ and $^3S_1$ two-nucleon states, respectively,
which read
\bea
P^{(^1S_0)}_a = \frac{1}{\sqrt8}\tau_2\tau_a\sigma_2\,,
\ \ \ 
P^{(^3S_1)}_i = \frac{1}{\sqrt8}\tau_2\sigma_2\sigma_i\,,
\eea
where $\vec{\sigma}$ and $\vec{\tau}$ are Pauli matrices
operating in the spin and isospin spaces, respectively.

\begin{figure}
\begin{center}
\epsfig{file=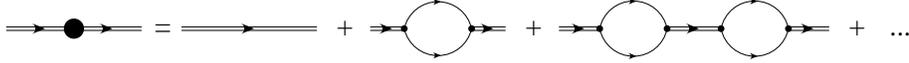,width=12cm}
\caption{\it Diagrams for dressed di-baryon propagator: 
a double line denotes a di-baryon field and a line does a nucleon field.
A double line with a filled circle denotes a dressed di-baryon 
field.
} 
\label{fig;dressed-dibaryon-propagator}
\end{center}
\end{figure}
Diagrams for dressed di-baryon propagator are 
shown in Fig.~\ref{fig;dressed-dibaryon-propagator}.
\footnote{Diagrams are prepared by using a C++ library, 
''FeynDiagram"\cite{feyndiagram}.}
As shown in Ref.~\cite{ksw-plb98}, two-nucleon bubble diagrams are
summed up to the infinite order, and thus
the inverse of dressed di-baryon propagator 
in the center of mass frame 
is obtained as
\bea
iD_{s,t}^{-1}(p) &=&  i \left[
\sigma_{s,t} \left(
E + 0 + \Delta_{s,t}
\right)
+ y_{s,t}^2 \frac{m_N}{4\pi}(\mu+ip)
\right]
\nnb \\
&=& iy_{s,t}^2 \frac{m_N}{4\pi}\left[
\frac{4\pi\sigma_{s,t} \Delta_{s,t}}{m_Ny_{s,t}^2} 
+ \frac{4\pi \sigma_{s,t}}{m_N^2y_{s,t}^2} p^2
+\mu
+ ip
\right]\,. 
\label{eq;dibaryon-propagator}
\eea
Energy of the nucleon in the center of mass frame reads 
$E=p^2/m_N$, and we have employed DR and PDS scheme
where $\mu$ is the renormalization scale
for the loop integration.

\begin{figure}
\begin{center}
\epsfig{file=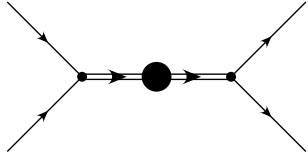,width=4cm}
\caption{\it Diagram for $S$-wave $NN$ scattering amplitude:
a double line with a filled circle is the dressed di-baryon propagator
obtained in Fig.~\ref{fig;dressed-dibaryon-propagator}. }
\label{fig;NNAmp}
\end{center}
\end{figure}

A diagram for the $S$-wave $NN$ scattering is 
given in Fig.~\ref{fig;NNAmp}.
Corresponding scattering amplitude is obtained as 
\bea
iA_{s,t} &=&
\frac{4\pi}{m_N}\frac{i}{
-\mu-\frac{4\pi\sigma_{s,t}\Delta_{s,t}}{m_Ny^2}
-\frac{4\pi\sigma_{s,t}}{m_N^2y_{s,t}^2}p^2 - i p}\,.
\label{eq;Ad}
\eea
On the other hand the $S$-wave amplitude in terms of the 
phase shift is given by
\bea
A  = \frac{4\pi}{m_N} \frac{1}{p\cot \delta_0 -ip}\,,
\label{eq;Aere}
\eea
where $p\cot \delta_0$ is expanded in terms of the effective range 
parameters as 
\bea
p \cot \delta_0 = -\frac{1}{a} + \frac12 rp^2 + \cdots\,.
\label{eq;pcotdel0}
\eea
$a$ is the scattering length, $r$ is the effective range, 
and the dots denote higher order terms.
With only di-baryon fields, Eq.~(\ref{eq;Ad}) gives
the corrections up to the quadratic term of $p$, i.e. $p^2$,
but if we include the one-pion-exchange contributions,
we 
have terms in higher orders of $p$. 
We note that the two nucleon bubble diagram plays a role similar   
to a kinetic term 
in mass renormalization in a propagator,
and the scattering amplitude is normalized by the term
proportional to $-ip$.
Consequently, with the term proportional to $1/y^2$ 
due to the normalization of the amplitude,
we can easily make linear relations 
to the effective range terms being expanded
in the denominator of the amplitude.
Equating Eq.~(\ref{eq;Ad}) with Eq.~(\ref{eq;Aere}), we have
\bea
-\frac1a = -\mu-\frac{4\pi\sigma_{s,t}\Delta_{s,t}}{m_Ny_{s,t}^2} ,
\ \ \ \ 
r = -\frac{8\pi\sigma_{s,t}}{m_N^2y_{s,t}^2}.
\eea

In general, Lagrangian for the di-baryons corresponding to 
the effective range parameters may be written 
as\footnote{A slightly different expression of the Lagrangian
for the deuteron pole has been obtained by Grie\ss hammer.
See Eq.~(2.1) in Ref.~\cite{g-npa04}. }
\bea
{\cal L}_d = 
d_a^\dagger 
\sum_{n=0}^\infty 
D_n 
\left[
i v\cdot \partial
+ \frac{(v\cdot \partial)^2-\partial^2}{4m_N}
\right]^n
d_a\,,
\eea
where $d=s,t$. The terms of the Lagrangian only require the Galilean 
invariance for two nucleons, and $D_n$ are coefficients, with the factor 
$1/y^2$ from the leading $dNN$ interaction, to match with the
effective range parameters.

\vskip 3mm \noindent 
{\bf 3. Di-baryons and perturbative pions}

Effective Lagrangian with di-baryon fields and pions 
may read
\bea
{\cal L} &=& {\cal L}_\pi + {\cal L}_{\pi N} 
+ {\cal L}_s
+ {\cal L}_t,
\eea
where ${\cal L}_\pi$ and ${\cal L}_{\pi N}$ are 
the chiral Lagrangians for pion and one-nucleon sector,
respectively.
Because we do not consider pion loops,
${\cal L}_s$ and ${\cal L}_t$ takes the same form as
given in Eqs.~(\ref{eq;Ls}) and (\ref{eq;Lt}), respectively.\footnote{
The chiral Lagrangians with di-baryon fields have 
been obtained by Soto and Tarr\'us~\cite{st-prc08,st-prc10}.
} 

Effective chiral Lagrangians ${\cal L}_\pi$ and ${\cal L}_{\pi N}$ 
are standard ones in the literature and we have
\bea
{\cal L}_\pi &=& 
- f_\pi^2{\rm Tr}(\Delta\cdot \Delta)
+ \frac{f_\pi^2}{4}{\rm Tr}(\chi_+)\,,
\\
{\cal L}_{\pi N} &=& 
N^\dagger \left[iv\cdot D-2ig_A S\cdot \Delta\right] N 
+ \frac{1}{2m_N}N^\dagger \left[
(v\cdot D)^2-D^2+\cdots \right] N,
\eea
with $D_\mu = \partial_\mu + \Gamma_\mu$
where $\Gamma_\mu = \frac12 [\xi^\dagger,\partial_\mu\xi]$,
$\Delta_\mu= \frac12\{\xi^\dagger,\partial_\mu \xi\}$,
and $\chi_+ = \xi^\dagger  \chi \xi^\dagger + \xi\chi^\dagger \xi $;
$\chi$ is the symmetry breaking term generating the pion mass,
$\chi \propto m_\pi^2$ at leading order,
and $\xi$ is the chiral field,
$\xi\xi=\exp(i\vec{\tau}\cdot\vec{\pi}/f_\pi)$.
$f_\pi$ is the pion decay constant
and $g_A$ is the axial vector coupling constant.
Dots denote terms irrelevant to this work.

\begin{figure}
\begin{center}
\epsfig{file=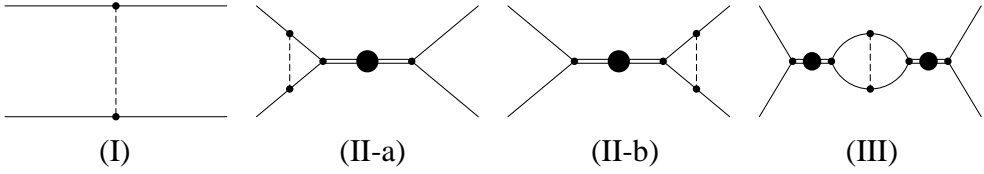,width=13cm}
\caption{\it Diagrams for contributions from one-pion-exchange
to the $NN$ scattering amplitude:
a dashed line denotes a ``potential'' pion field. See the caption 
in Fig.~\ref{fig;dressed-dibaryon-propagator} as well. }
\label{fig;one-pion-exchange-diagrams}
\end{center}
\end{figure}
With the one-pion-exchange contributions 
(so called ``potential'' pion) at ${\cal O}(Q^0)$ shown
in Fig.~\ref{fig;one-pion-exchange-diagrams},
we may write down the whole amplitude as
\bea
A = A_d + A_\pi\,,
\eea
where $A_d$ is the amplitude from the tree-level dressed di-baryon 
contributions without pions, $A_{s,t}$ in Eq.~(\ref{eq;Ad}), and
$A_\pi$ is the amplitude from the one-pion-exchange contributions
in Fig.~\ref{fig;one-pion-exchange-diagrams}. 
We may divide it as
\bea
A_\pi = A_{(I)} + A_{(II)} + A_{(III)}\,,
\eea 
where the amplitudes, $A_{(I)}$, $A_{(II)}$, 
and $A_{(III)}$ 
are obtained 
as 
\bea
A_{(I)} &=& \left(
\frac{g_A}{2f_\pi}
\right)^2 \left[
-1
+\frac{m_\pi^2}{4p^2}\ln\left(
1
+ \frac{4p^2}{m_\pi^2}
\right)
\right]\,,
\label{eq;api1}
\\
A_{(II)} &=&
\frac{1}{O_d-ip}
\left(\frac{g_A}{2f_\pi}\right)^2 \left\{
-2\mu
+ \frac{m_\pi^2}{p} \arctan\left(
\frac{2p}{m_\pi}\right)
+2ip\left[
-1
+ \frac{m_\pi^2}{4p^2}\ln\left(
1
+ \frac{4p^2}{m_\pi^2}
\right)
\right]
\right\}\,,
\label{eq;api2}
\nnb \\
\\
A_{(III)} &=&
\frac{1}{(O_d-ip)^2}\left(
\frac{g_A}{2f_\pi}\right)^2\left\{
-\mu^2 + m_\pi^2 
+m_\pi^2 \ln\left(\frac{\mu}{m_\pi}\right)
-p^2
\right. \nnb \\ && \left.
+2ip\left[
-\mu
+ \frac{m_\pi^2}{2p}\arctan\left(
\frac{2p}{m_\pi}
\right) 
\right]
-2p^2\left[
-1
+\frac{m_\pi^2}{4p^2}\ln\left(
1 + \frac{4p^2}{m_\pi^2}
\right)
\right]
\right\}\,,
\label{eq;api3}
\eea
with 
\bea
O_d = -\mu 
-\frac{4\pi\sigma\Delta}{m_Ny^2}
-\frac{4\pi\sigma}{m_N^2y^2}p^2\,.
\eea 
The subscripts $s,t$ are dropped from the di-baryon 
parameters, $\sigma$, $\Delta$, and $y$ for simplicity.
The loop diagrams have been calculated by using DR
and PDS scheme, as done in Refs.~\cite{ksw-plb98,ksw-npb98}.

Expanding the one-pion-exchange amplitude
$A_\pi$ around the inverse of the di-baryon amplitude $1/A_d$,
which is associated with the non-trivial fixed point,
we have
\bea
p\cot\delta_0 &=& 
ip + \frac{4\pi}{m_N}\frac{1}{A}
=
ip + \frac{4\pi}{m_N}\frac{1}{A_d+A_\pi}
\simeq
ip + \frac{4\pi}{m_N}\frac{1}{A_d}
-\frac{4\pi}{m_N}\frac{A_\pi}{A_d^2}\,.
\label{eq;pcot21}
\eea
We note that because the one-pion-exchange correction 
is obtained in the form of
$A_\pi/A_d^2$ in the above expression, 
the pole structures 
due to the di-baryon propagator 
in $A_{(II)}$ and $A_{(III)}$
disappear, whereas $A_d^{-2}$ and $A_d^{-1}$
terms appear as interactions
($O_d^2$ and $O_d$) along with $A_{(I)}$ and $A_{(II)}$ terms,
respectively. 

Inserting Eqs.~(\ref{eq;Ad}) and (\ref{eq;api1}-\ref{eq;api3})
into Eq.~(\ref{eq;pcot21}), we obtain
\bea
\lefteqn{
p\cot \delta_0 =
O_d
-\frac{g_A^2m_N}{16\pi f_\pi^2}\left\{
-\mu^2 
+ m_\pi^2
+m_\pi^2\ln\left(\frac{\mu}{m_\pi}\right)
-p^2
\right.
} \nnb \\ && \left.
+2O_d\left[
-\mu
+\frac{m_\pi^2}{2p}
\arctan\left(\frac{2p}{m_\pi}\right)
\right]
+(O_d^2-p^2)\left[
-1
+\frac{m_\pi^2}{4p^2}\ln\left(
1
+\frac{4p^2}{m_\pi^2}
\right)
\right]
\right\}\,.
\label{eq;pcotdel0-pions}
\eea
We note that the right hand side of Eq.~(\ref{eq;pcotdel0}) is real,
and Eq.~(\ref{eq;pcotdel0-pions}) satisfies the unitary condition 
for the phase shift $\delta_0$:
the phase shift $\delta_0$ should be real below the threshold
of pion production. 
Satisfying this condition is not so trivial,
as recently pointed out in Ref.~\cite{ly-11}.
For example,
in the scheme proposed by Kaplan, Savage and Wise \cite{ksw-plb98}, 
one performs resummation of the leading contact interaction $C_0$, 
and the next-leading contact interaction $C_2$ is 
treated perturbatively in calculation of the amplitude. 
The unitarity would be satisfied order by order in the 
perturbative expansion.
This treatment makes the phase shift $\delta_0$ complex 
in the higher order corrections, and thus one needs an additional 
expansion of $\tan\delta_0$ in terms of the phase shift $\delta_0$.

\vskip 3mm \noindent
{\bf 4. Scattering length and effective range: renormalization}

We renormalize our results with two effective range parameters,
scattering length and effective range in the expression, 
\bea
p \cot\delta_0 &=& -\frac{1}{a} + \frac12 rp^2 + F(p)\,,
\label{eq;pcotdelF}
\eea
where $F(p)$ is a function for effective range parameters in 
higher orders of $p$; we can approximate $F(p)\propto p^4$ 
when $p\ll 1$. 

If we define   $a_d(\mu)$ and $r_d(\mu)$ as
\bea
\frac{1}{a_d(\mu)} 
\equiv 
\left\{
\frac{1}{a} 
- \frac{g_A^2m_N}{16\pi f_\pi^2}\left[
-\mu^2
+m_\pi^2
+m_\pi^2\ln\left(\frac{\mu}{m_\pi}
\right)
\right]
\right\}
\left[ 1 -\frac{g_A^2m_N}{8\pi f_\pi^2}(-\mu+m_\pi)
\right]^{-1},
\label{eq;1/ad(mu)}
\\
\frac12r_d(\mu) 
\equiv 
\left\{
\frac12 r
+ \frac{g_A^2m_N}{16\pi f_\pi^2}\left[
-1
+\frac83\frac{1}{m_\pi a_d(\mu)}
-\frac{2}{m_\pi^2 a_d^2(\mu)}
\right]
\right\} 
\left[ 1 -\frac{g_A^2m_N}{8\pi f_\pi^2}(-\mu+m_\pi)
\right]^{-1},
\eea
then we obtain the higher order contributions as
\bea
F(p) &=& - \frac{g_A^2m_N}{16\pi f_\pi^2}\left\{
\left[
-\frac83\frac{1}{m_\pi a_d(\mu)}
+ \frac{2}{m_\pi^2 a_d^2(\mu)}
\right]p^2
\right. \nnb \\ && 
+ 2\left( 
-\frac{1}{a_d(\mu)}
+\frac12 r_d(\mu)p^2
\right)\left[
-m_\pi
+ \frac{m_\pi^2}{2p}\arctan\left(
\frac{2p}{m_\pi}
\right)
\right]
\nnb \\ && \left.
+\left[\left(
-\frac{1}{a_d(\mu)}
+ \frac12r_d(\mu)p^2
\right)^2
-p^2
\right]\left[
-1
+ \frac{m_\pi^2}{4p^2}\ln\left(
1
+ \frac{4p^2}{m_\pi^2}
\right)
\right]
\right\}\,.
\label{eq;Fp}
\eea
For the renormalization of the scattering length $a$
and the effective range $r$, we use the empirical 
values of the parameters as 
\bea
\frac{1}{a_0} = -8.30 \ \mbox{\rm (MeV)}\,,
\ \ \ \
r_0 = 2.73 \ \mbox{\rm (fm)}\,,
\eea
for the $^1S_0$ channel and 
\bea
\frac{1}{a_1} = 36.4 \ \mbox{\rm (MeV)}\,,
\ \ \ \ 
r_1 = 1.76 \ \mbox{\rm (fm)}\,,
\eea
for the $^3S_1$ channel\cite{nnonline}. 
%
Conventionally, as discussed in the introduction,
the terms, $1/a_d(\mu)$ and $r_d(\mu)$, in the function $F(p)$
are replaced by the physical quantities, $1/a$ and $r$, respectively.
In this work, we rather retain 
the $1/a_d(\mu)$ and $r_d(\mu)$ terms in the higher orders
as a probe of the optimal high energy and higher order contributions,
and study the $\mu$-dependence in the higher order terms,
the effective range corrections, $v_2$, $v_3$, and $v_4$,
and the S-wave phase shifts in higher momentum regions below.

The value of the scale parameter $\mu$ has been chosen as
$\mu=m_\pi$ \cite{ksw-plb98,ksw-npb98}.
With this particular choice of the scale parameter $\mu$, 
we have
\bea
\frac{1}{a_d(\mu)} = \frac{1}{a_0}\,\,\,
\mbox{and}\,\,\, \frac{1}{a_1}\,,
\eea
for the $^1S_0$ and $^3S_1$ channel, respectively.
This implies that there are no contributions to $a_d(\mu)$
from the one-pion-exchange corrections,
and it is determined solely by the 
di-baryon contribution. 
It was shown, e.g., in Ref.~\cite{hyun2000} that if the value
of a scale parameter is chosen around the mass of the lightest 
degree of freedom that is integrated out, then the results for the
phase shifts in the $^1 S_0$ channel agree very well with the 
experimental data.
Because we consider only the one-pion-exchange in this work, 
we vary the scale parameter $\mu$
up to $\sim 400$ MeV.

\begin{figure}
\begin{center}
\epsfig{file=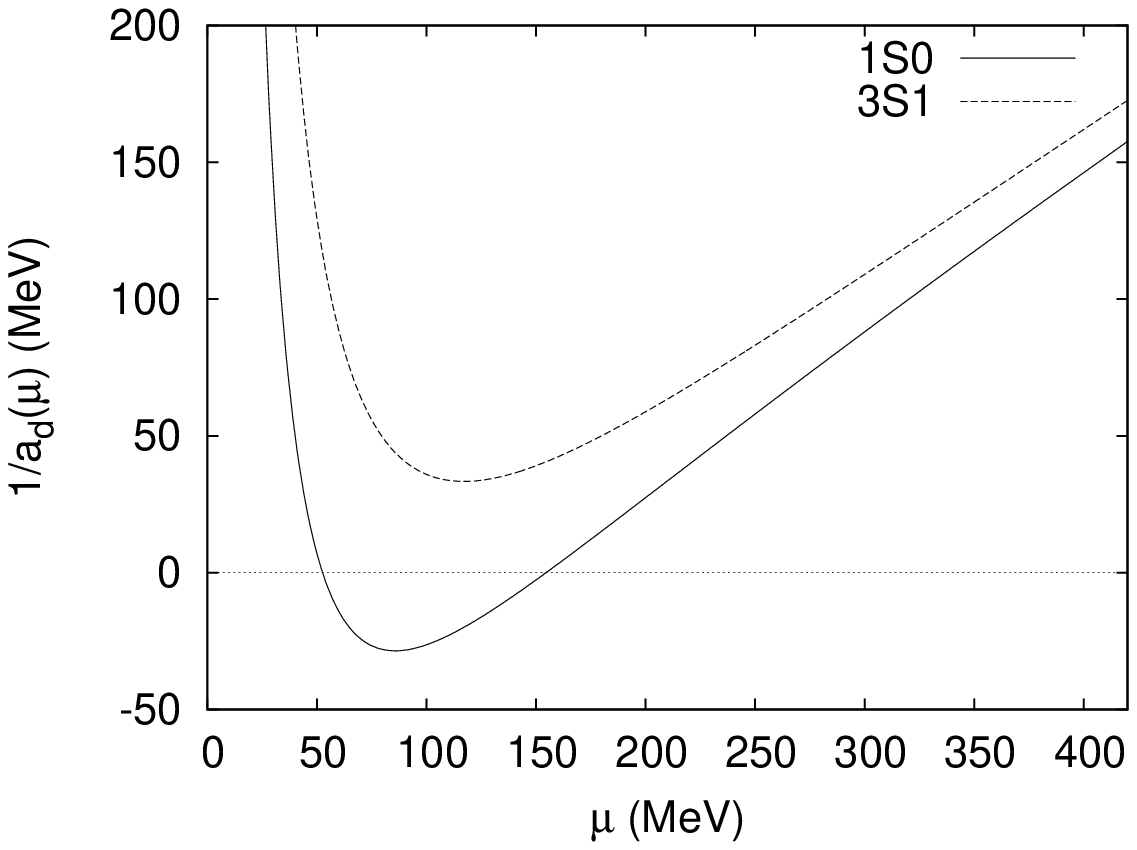,width=7cm}
\epsfig{file=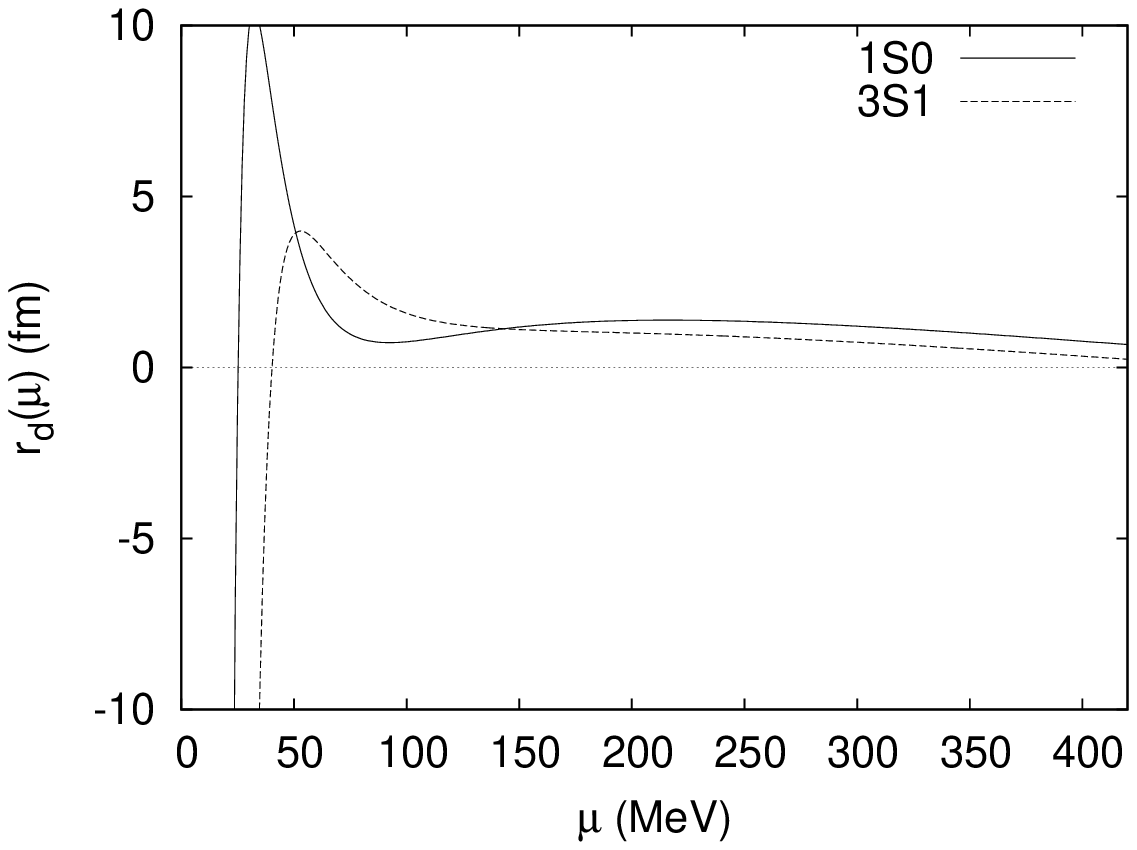,width=7cm}
\caption{\it $1/a_d(\mu)$ and $r_d(\mu)$ 
for $^1S_0$ and $^3S_1$ channels are
plotted as functions of the scale parameter $\mu$.}
\label{fig;invad-rd2}
\end{center}
\end{figure}

We plot $1/a_d(\mu)$ and $r_d(\mu)$ as functions of $\mu$ in
Fig.~\ref{fig;invad-rd2}.
We can see that $1/a_d(\mu)$ vanishes for the $^1S_0$ channel at 
$\mu = 52.6$ and 154.6 MeV, whereas there is no point where $1/a_d(\mu)$
vanishes for the $^3S_1$ channel.

For physically meaningful values of $\mu$, i.e. 
$\mu > m_\pi$, the value of $1/a_d(\mu)$ increases monotonically
as $\mu$ increases.
At sufficiently large $\mu$, we can approximate Eq.~(\ref{eq;1/ad(mu)})
as 
\begin{equation}
\frac{1}{a_d (\mu)} \simeq 
\frac{1}{2} \mu,
\end{equation}
and thus $1/a_d(\mu)$ increases linearly with respect to $\mu$. 
For large $\mu$, we can approximate $r_d(\mu)$ as
\begin{equation}
r_d(\mu) \simeq 
\frac{1}{2 \mu} \left(
-1 + \frac{4\mu}{3 m_\pi} - \frac{\mu^2}{2 m_\pi^2} \right).
\end{equation}
Because $\mu$'s in the parenthesis are divided by $m_\pi$
and we have $\mu^{-1}$ in front of the parenthesis, 
we have a relatively slow change of
$r_d(\mu)$, and it eventually decreases because of the quadratic
term in the parenthesis.

\begin{figure}
\begin{center}
\epsfig{file=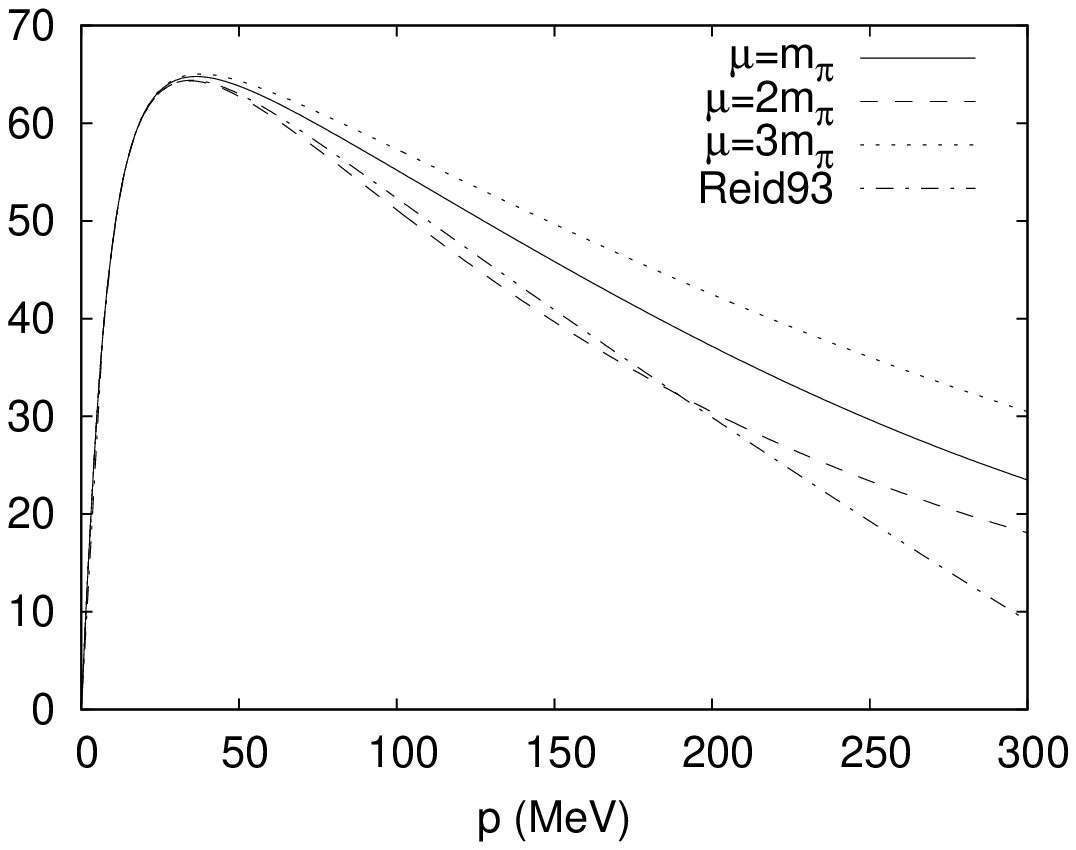,width=7cm}
\epsfig{file=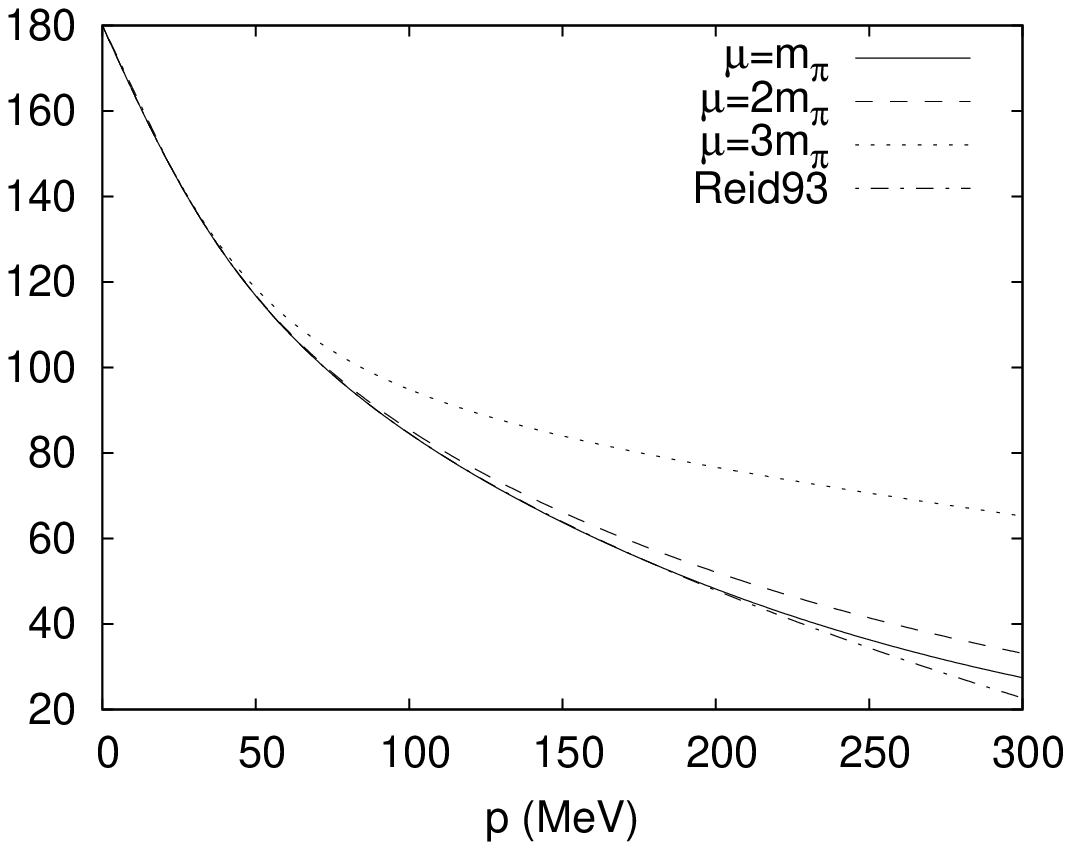,width=7cm}
\caption{\it $S$-wave phase shifts for $^1S_0$ channel (left panel)
and for $^3S_1$ channel (right panel):
Curves of 
our results with $\mu = m_\pi$, $2m_\pi$, $3m_\pi$
are labeled by the $\mu$ values,
and those labeled by "Reid93" are 
from a modern phenomenological potential.}
\label{fig;cot-deft-nmpi}
\end{center}
\end{figure} 

Fig.~\ref{fig;cot-deft-nmpi} shows the phase shifts in the
$^1 S_0$ (left) and $^3 S_1$ (right) channels obtained from
Eq.~(\ref{eq;pcotdelF}).
If one requires that a result from EFT should be insensitive
to the choice of a value of cutoff, 
corresponding to the scale parameter $\mu$ in this work,
the applicable range of the present calculation for the 
$S$-wave phase shifts is quite restricted, less than $p\sim  50$ MeV
when we choose the scale parameter $\mu$ as $\mu=m_\pi$, 2$m_\pi$,
and 3$m_\pi$, as shown in Fig.~\ref{fig;cot-deft-nmpi}.
However since we integrate out the degrees of freedom heavier than
the one-pion exchange, it may be reasonable to consider 
the scale parameter in the range $m_\pi < \mu < 2 m_\pi$.
Within this range, $^1 S_0$ channel shows the best results with
the $\mu$ close to $2 m_\pi$. This behavior is consistent with
the one observed in Ref.~\cite{hyun2000}.
For the $^3 S_1$ channel the results with reasonable range of
$\mu$ show good agreement with each other to substantially
large momenta.

It is worthwhile to consider the radius of 
convergence for which the perturbative expansion of 
Eq.~(\ref{eq;pcot21}) is reasonable. Since the scattering length
and effective range terms are determined with 
the empirical values,
the term $F(p)$ includes the
contributions from one-pion exchange and 
depends on $\mu$.
We consider the ratio defined by
\begin{equation}
R \equiv \frac{|F(p)|}{-1/a + r p^2/2}
\label{eq;ratio}
\end{equation}
as a criterion for the convergence.

\begin{figure}
\begin{center}
\epsfig{file=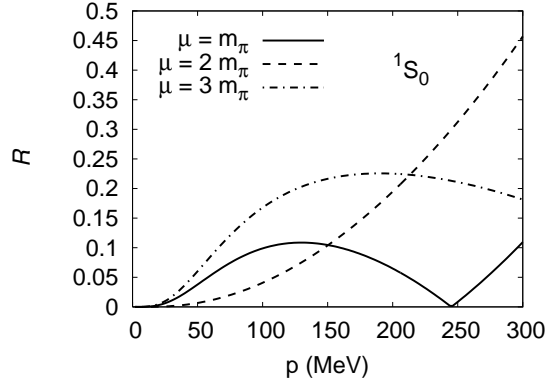, width= 7.5cm}
\end{center}
\caption{\it Ratio $R$ in Eq.~(\ref{eq;ratio})
as a function of momentum $p$ with various values of $\mu$.}
\label{fig;ratio}
\end{figure}

Fig.~\ref{fig;ratio} shows the ratio $R$ in the $^1 S_0$ channel
with $\mu = m_\pi$, $2 m_\pi$ and $3 m_\pi$.
Dependence on the $\mu$ value is significant,
but regardless of the $\mu$ value, one can say that the perturbative
expansion of the pion-exchange contribution is reasonable for
$p \leq 150$ MeV.
Agreement of the phase shifts to the phenomenological ones is
good within this range.

\vskip 3mm \noindent
{\bf 5. Higher order effective range corrections}

After renormalization of the scattering length and the effective 
range, the function $F(p)$ in Eq.~(\ref{eq;Fp})
has no unknown parameters, and depends only on the scale parameter $\mu$.  
Expanding $p \cot \delta_0$ to higher orders, 
one has
\bea
p\,\cot \delta_0 = -\frac{1}{a} 
+ \frac12 r p^2
+ v_2 p^4
+ v_3 p^6
+ v_4 p^8
+ \cdots\,.
\eea
We renormalize the scattering length $a$ and 
the effective range $r$ to the physical values,
and obtain the higher order parameters,
$v_2$, $v_3$, and $v_4$, ... 
by expanding $F(p)$ in Eq.~(\ref{eq;Fp}) in powers of $p$ as
\bea
F(p) = v_2 p^4 + v_3 p^6 + v_4 p^8 + \cdots.
\eea
We obtain
\bea
v_2 &=& \frac{g_A^2m_N}{16\pi f_\pi^2}\left\{
-\frac{16}{3} 
\frac{1}{a_d^2(\mu)m_\pi^4}
+ \frac{32}{5} 
\frac{1}{a_d(\mu)m_\pi^3}
- \frac{2}{m_\pi^2}\left[1+\frac{r_d(\mu)}{a_d(\mu)}\right]
+ \frac43 \frac{r_d(\mu)}{m_\pi}
\right\}\,, 
\\
v_3 &=& \frac{g_A^2m_N}{16\pi f_\pi^2}\left\{
16 \frac{1}{a_d^2(\mu)m_\pi^6}
- \frac{128}{7}
\frac{1}{a_d(\mu)m_\pi^5}
+ \frac{16}{3}\frac{1}{m_\pi^4}
\left[ 1 + \frac{r_d(\mu)}{a_d(\mu)}\right]
\right. \nnb \\ && \left.
- \frac{16}{5}\frac{r_d(\mu)}{m_\pi^3}
+\frac12\frac{r_d^2(\mu)}{m_\pi^2}
\right\}\,, 
\\
v_4 &=& \frac{g_A^2m_N}{16\pi f_\pi^2}\left\{
-\frac{256}{5}\frac{1}{a_d^2(\mu)m_\pi^8}
+\frac{512}{9}\frac{1}{a_d(\mu)m_\pi^7}
-16\frac{1}{m_\pi^6}\left[
1 + \frac{r_d(\mu)}{a_d(\mu)}\right]
\right. \nnb \\ && \left.
+\frac{64}{7}\frac{r_d(\mu)}{m_\pi^5}
-\frac43\frac{r_d^2(\mu)}{m_\pi^4}
\right\}\,. 
\eea
We note that the expressions of $v_2$, $v_3$ and $v_4$ above are 
obtained in terms of $1/a_d(\mu)$ and $r_d(\mu)$. 
Those expressions have been obtained 
by Cohen and Hassen without $r_d(\mu)$ and in terms
of the physical scattering length $a$~\cite{ch1-prc99,ch2-prc99}. 
We can reproduce their results by choosing $\mu=m_\pi$
and neglecting the contribution $r_d(\mu)$.

Values of $v_2$, $v_3$, and $v_4$ are shown in Table \ref{table;v234}.
\begin{table}
\begin{center}
\begin{tabular}{c|c|ccc} \hline
 & & $v_2$(fm$^3$) & $v_3$(fm$^5$) & $v_4$(fm$^7$) \\ \hline
$^1S_0$
 & PWA & $-$0.48 & 3.8 & $-$17.0 \\ 
 & $a^{phy.}$, $r=0$ & $-$3.32 & 17.8 & $-$106.8 \\
 & $a^{phy.}$, $r^{phy.}$ & 0.69 & 3.99 & $-$25.9 \\
 & $\mu=m_\pi$ & $-$1.74 & 11.1 & $-$68.0 \\
 & $\mu=2m_\pi$ & 0.17 & 0.71 & $-$4.85 \\
 & $\mu=3m_\pi$ & $-$2.75 & 17.8 & $-$117 \\
 & $\mu=178$ MeV & $-$0.48$^*$ & 4.40 & $-$26.8 \\
 & $\mu=330$ MeV & $-$0.48$^*$ & 4.37 & $-$29.0 \\
 & FMS(NLO)\cite{fms-npa00} & $-$3.3 & 19 & $-$117 \\
 & FMS(NNLO)\cite{fms-npa00} & $-$1.2 & 2.9 & $-$0.7 \\
\hline
$^3S_1$ 
 & PWA & 0.04 & 0.67 & $-$4.0 \\ 
 & $a^{phy.}$, $r=0$ & $-$0.96 & 4.57 & $-$25.5 \\
 & $a^{phy.}$, $r^{phy.}$ & 0.44 & 0.48 & $-$2.88 \\
 & $\mu=m_\pi$ & $-$0.05 & 1.42 & $-$8.22 \\
 & $\mu=2m_\pi$ & $-$0.25 & 2.39 & $-$16.2 \\
 & $\mu=3m_\pi$ & $-$3.23 & 21.2 & $-$139 \\
 & $\mu=167$ MeV & 0.04$^*$ & 0.75 & $-$4.1 \\
 & $\mu=246$ MeV & 0.04$^*$ & 0.68 & $-$4.7 \\
\hline
\end{tabular}
\caption{\it Values of $v_2$, $v_3$, and $v_4$ for
$^1S_0$ and $^3S_1$ channel:
values in the first row for each of the channels  
are from partial wave analysis (PWA),
those in the second row are obtained by putting $a_d(\mu)=a^{phys.}$ and 
$r_d(\mu)=0$ in our results, 
those in the third row by putting $a_d(\mu)=a^{phys.}$ and
$r_d(\mu)=r^{phys.}$, 
and those from the fourth to the eighth row are our results
with different $\mu$ values: for the fourth to the sixth row 
$\mu=m_\pi$, $2m_\pi$, and $3m_\pi$, respectively,
and those in the seventh and eighth row, $\mu$ value is fixed 
by using the value of $v_2$ of PWA (the values with * above are fitted ones);
$\mu=$ 178 and 330 MeV for $^1S_0$ channel and $\mu=$ 167 and 246 MeV
for $^3S_1$ channel.
Values in the ninth and tenth rows for $^1S_0$ channel 
are results obtained by Fleming, Mehen, Stewart (FMS)~\cite{fms-npa00} 
up to NLO and NNLO in the KSW counting, respectively.
\label{table;v234}
}
\end{center}
\end{table}
The values in the first row for each channel are obtained 
from partial wave analysis (PWA).
\footnote{The values of PWA are taken
from Table I in Ref.~\cite{ch2-prc99}.}
Those in the second row are obtained by putting 
$a_d(\mu)=a^{phy.}$ and $r_d(\mu)=0$ where $a^{phy.}$ is the physical
value of the scattering length, $a^{phy.}=a_0$ for the $^1S_0$ channel
and $a^{phy.}=a_1$ for the $^3S_1$ channel. 
These results are equal to what Cohen and Hansen have 
obtained~\cite{ch2-prc99}.
Those in the third row are obtained by putting 
$a_d(\mu)=a^{phy.}$ and $r_d(\mu)=r^{phy.}$ where $r^{phy.}$ is the 
physical effective range, $r^{phy.}= r_0$ for the $^1S_0$ channel and 
$r^{phy.}=r_1$ for the $^3S_1$ channel.
We find that the corrections from the effective range are significant,
but it is not enough to reproduce the PWA values of 
$v_2$, $v_3$ and $v_4$. 
Those from the fourth to the eighth row are our results
with various $\mu$ values,
$\mu=m_\pi$, $2m_\pi$, $3m_\pi$, 
whereas 
$\mu = $ 178 and 330 MeV for $^1S_0$ channel
and $\mu=$ 167 and 246 MeV for $^3S_1$ channel are fitted
so that the values of $v_2$ of PWA are reproduced 
(we find several points to reproduce the value of $v_2$ of PWA
as to be seen below).
It has been recently discussed that the scale parameter 
may be considered as an additional short-range 
``counter term'' provided one allows for 
a fine tuning of the scale value~\cite{eg-epja09}. 
The results with $\mu=m_\pi$, $2m_\pi$ and $3m_\pi$ show a
sensitivity of $v_2$, $v_3$ and $v_4$ to the $\mu$ value, whereas 
the results fitted to the $v_2$ value of PWA 
fairly reproduce all of the values of $v_2$, $v_3$ and
$v_4$ with the adjusted values of the scale parameter $\mu$. 

Values of $v_2$, $v_3$, and $v_4$ in the ninth and tenth rows 
for the $^1S_0$ channel in Table \ref{table;v234} are obtained
by Fleming, Mehen, and Stewart (FMS)~\cite{fms-npa00} up to 
next-to leading order (NLO) and next-to-next-to leading order (NNLO)
in the KSW counting scheme, respectively.
They employed so called global fitting method 
where they used the phase shift data in the ranges, 
$p=7\sim$ 80~MeV for NLO 
and $p=7\sim$ 200~MeV for NNLO, in fitting the parameters.
In addition, they obtained slightly different effective 
ranges, $r=2.65$~fm for NLO and $r=2.63$~fm for NNLO
from the empirical value $r=2.73$~fm.
Their NLO results basically agree with those obtained by 
Cohen and Hansen~\cite{ch1-prc99,ch2-prc99}, whereas the 
NNLO results are improved but 
not adequate to completely reproduce the results of PWA.  
It may indicate the slow convergence of the higher order
corrections as discussed by the authors~\cite{fms-npa00}.
It is also noted that adjusting the $\mu$-value 
which improves the estimated values of $v_2$, $v_3$, and $v_4$ 
in this work
seems to play a similar role to the higher order 
corrections.

\begin{figure}
\begin{center}
\epsfig{file=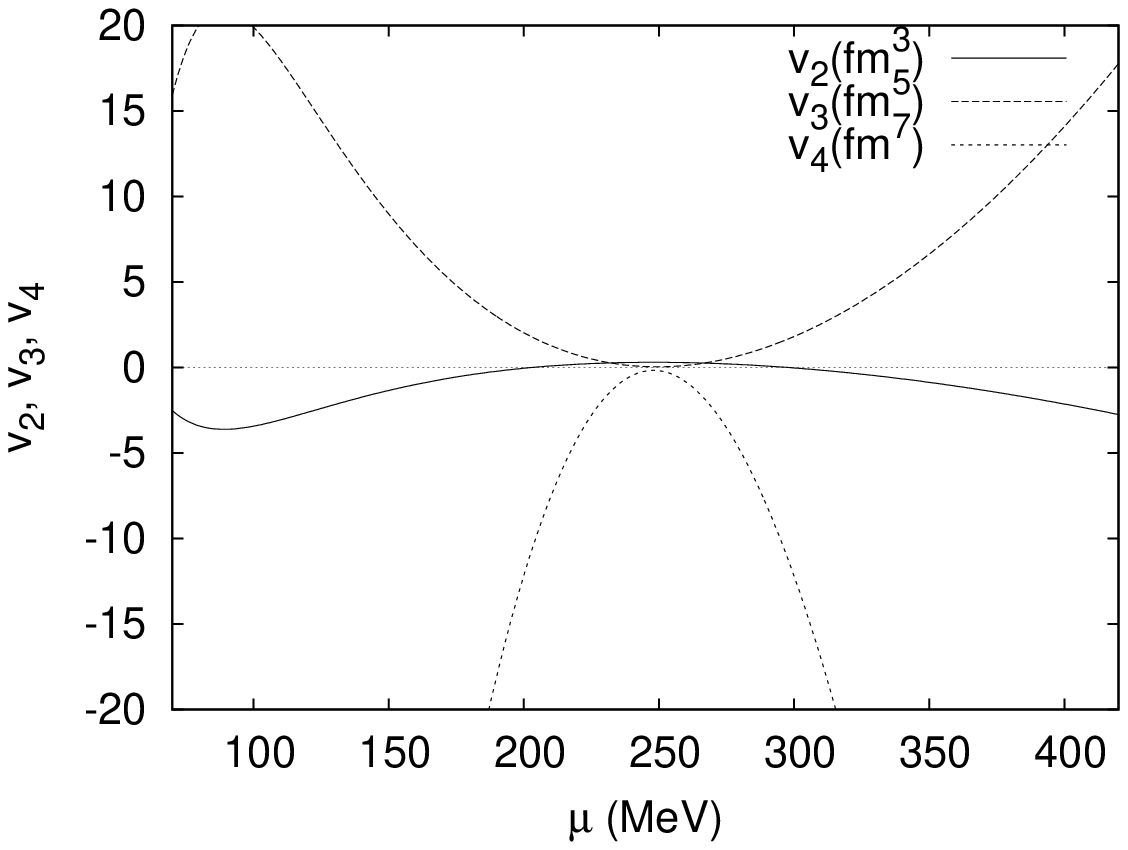,width=7cm}
\epsfig{file=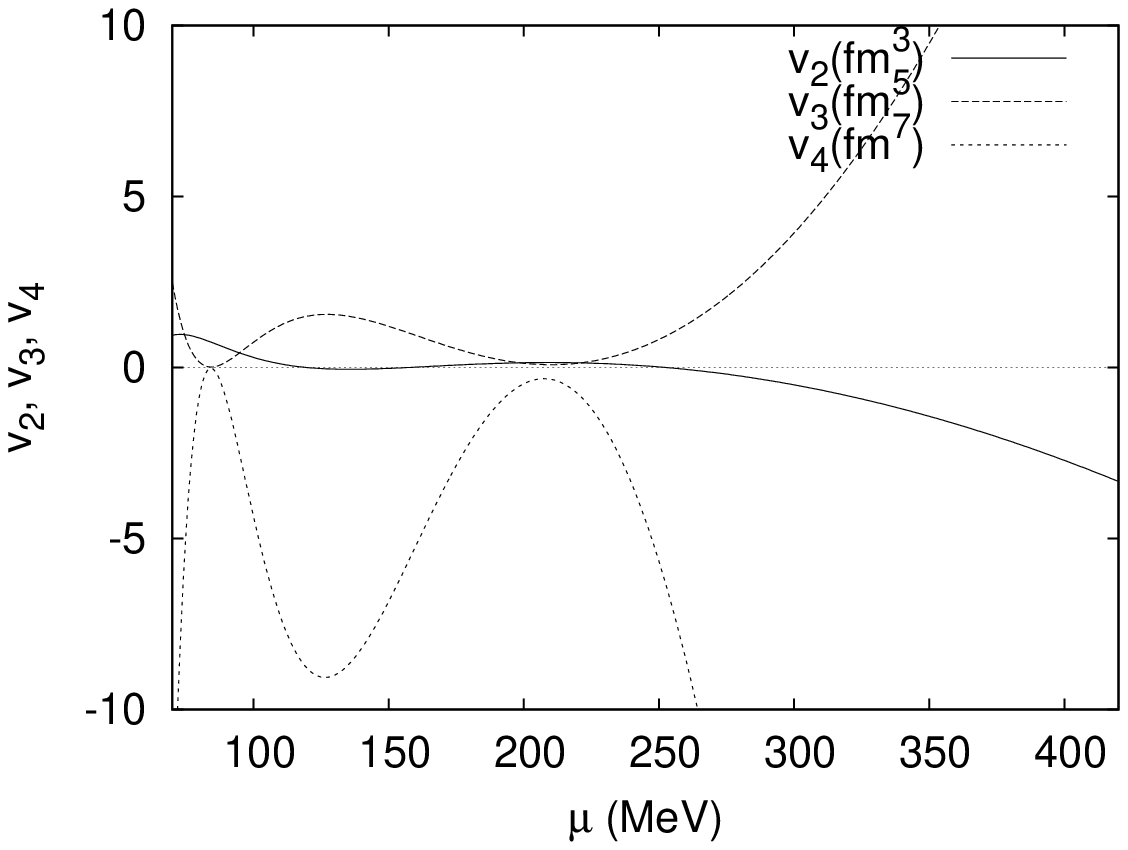,width=7cm}
\caption{\it
Effective range expansion parameters
$v_2$, $v_3$, and $v_4$ for $^1S_0$ channel (left panel) 
and $^3S_1$ channel (right panel) are plotted as functions
of the scale parameter $\mu$. 
\label{fig;v234}
}
\end{center}
\end{figure}

Now we plot our results of $v_2$, $v_3$, and $v_4$
for $^1S_0$ and $^3S_1$ channels
as functions of the scale parameter $\mu$ in Fig.~\ref{fig;v234}.  
We find the values of $v_2$, $v_3$, and $v_4$ are 
quite sensitive to the value of $\mu$:
the sensitivity increases as the order of the terms
increases with the units we have chosen,
$fm^3$, $fm^5$, $fm^7$, respectively.
Thus $v_4$, is the most sensitive, 
$v_3$ is in middle, and 
$v_2$ is less sensitive to $\mu$ than others.
    
\begin{figure}
\begin{center}
\epsfig{file=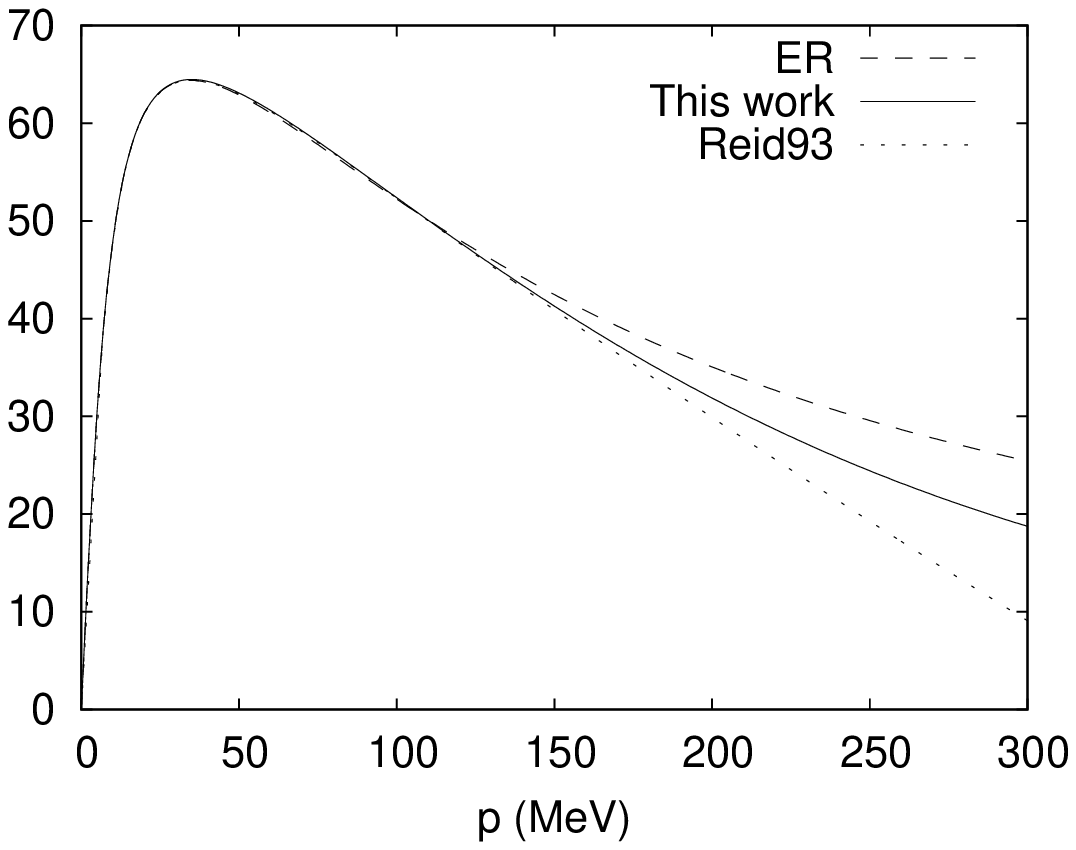,width=7cm}
\epsfig{file=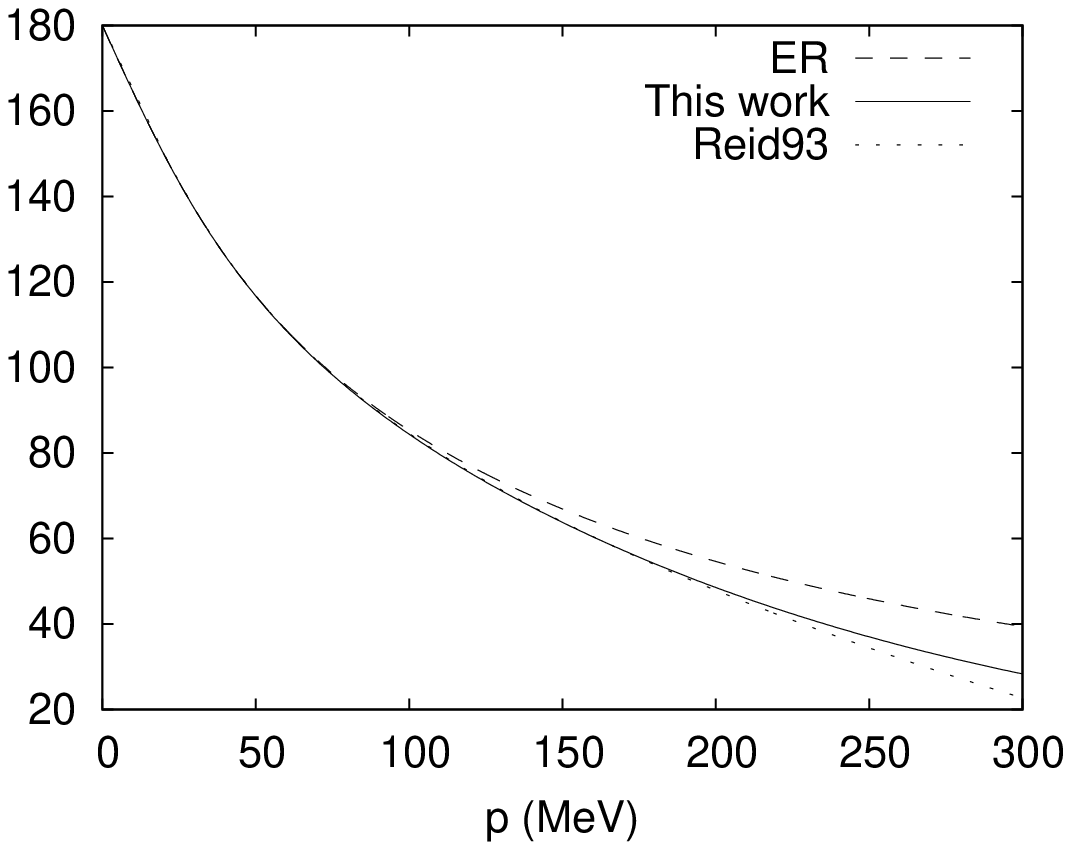,width=7cm}
\caption{\it $S$-wave phase shifts for $^1S_0$ channel (left panel)
and for $^3S_1$ channel (right panel):
Curves labeled by ``ER" are obtained from effective range expansion
with the first two coefficients, scattering length and effective range, 
those by "This work" are our results using $\mu = 178$ MeV for $^1S_0$ 
and $\mu= 167$ MeV for $^3S_1$ channel, and those by "Reid93" are 
from a modern phenomenological potential.}
\label{fig;cot-deft}
\end{center}
\end{figure}

The phase shifts for the $S$-waves are plotted 
in Fig.~\ref{fig;cot-deft} with Eq.~(\ref{eq;pcotdelF})
and the $\mu$ adjusted to PWA $v_2$ values. 
Our results (labeled by ``This work") 
are plotted with $\mu=178$ MeV for the $^1S_0$ channel
and $\mu=167$ MeV for the $^3S_1$ channel.
We compare our results with the ones which are obtained by
keeping only two effective range parameters, $a$ and $r$
in the effective range (ER) expansion,
and those from a modern phenomenological potential (Reid93) \cite{nnonline}.
Compared to the curve of ER, the results with one-pion exchange
are improved in both accuracy and the range where the theory
is applicable.
Comparing the results with the fine-tuned $\mu$ values
to those with randomly chosen $\mu$'s in Fig.~\ref{fig;cot-deft-nmpi},
we again see that both range and accuracy are improved.
It is notable that the fine-tuned values of $\mu$ are within the
range $m_\pi < \mu < 2 m_\pi$, which 
may support 
the perturbative treatment of pion-exchange interactions 
in the framework of di-baryon pionless EFT.

\begin{figure}
\begin{center}
\epsfig{file=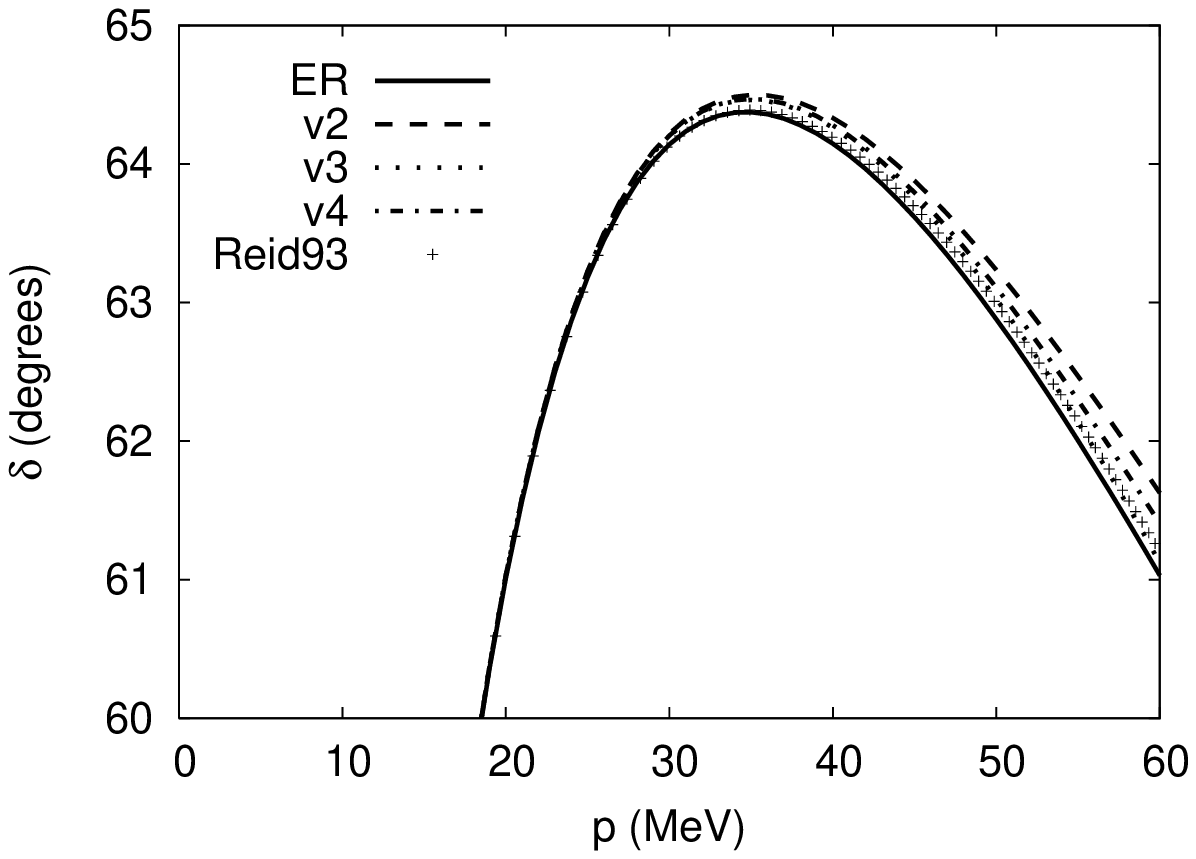, width=7cm}
\epsfig{file=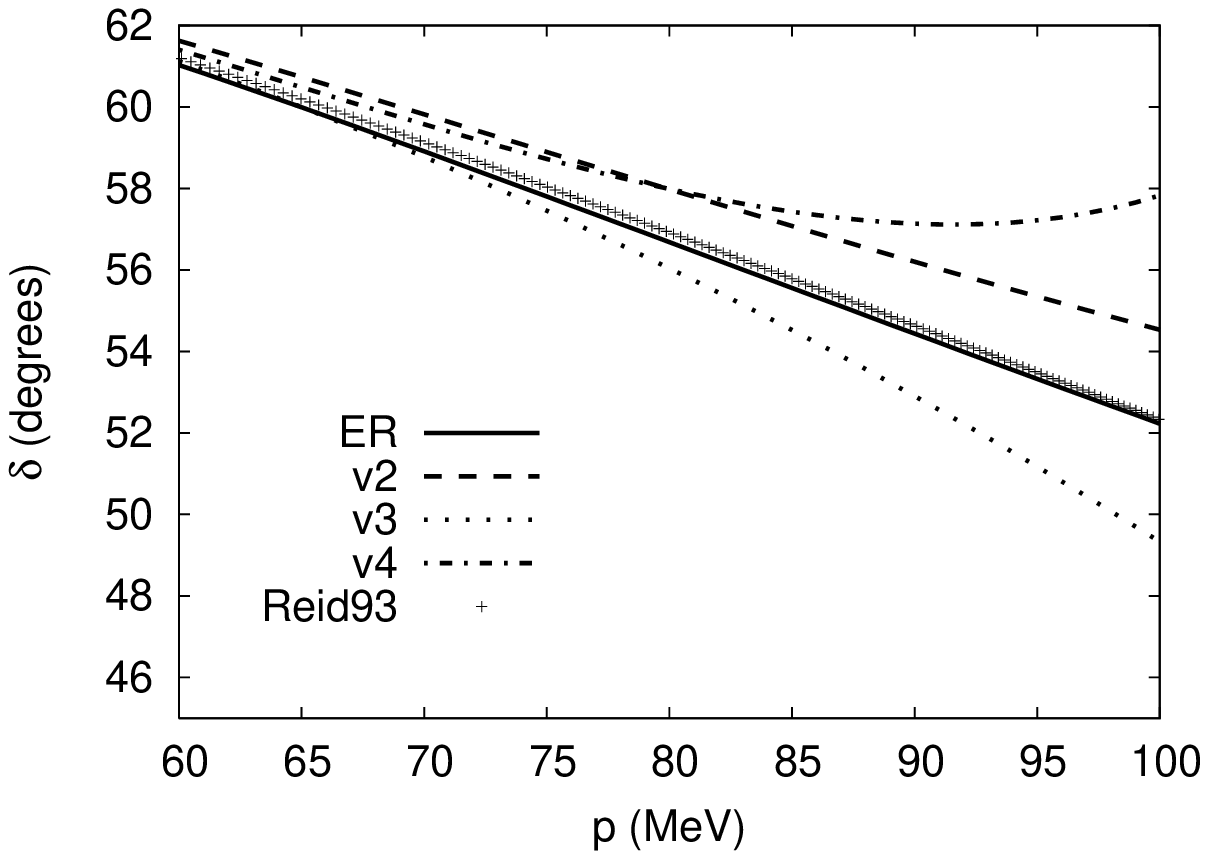, width=7cm}
\end{center}
\caption{\it Phase shifts in the $^1S_0$ channel for ER, ER$+v_2$,
ER$+v_2+v_3$ and ER$+v_2+v_3+v_4$. Left panel shows the results
for $0 <p<60$ MeV and the right for $60<p<100$ MeV.}
\label{fig;lowhigh}
\end{figure}

Finally we investigate the contributions of $v_2$, $v_3$, and $v_4$
terms to the phase shifts.
In Fig.~\ref{fig;lowhigh} we plot the phase shifts in the $^1 S_0$
channel for ER, and $v_2$, $v_3$ and $v_4$ terms added successively
with $\mu =178$ MeV. The results are compared with those of Reid93.
For the momenta $0 < p < 30$ MeV, the results of $v_2$, $v_3$
and $v_4$ almost coincide.
At around $p\sim 50$ MeV, the 
variation from $v_2$ to $v_3$ is
about half of 
that from ER to $v_2$, and the 
variation from
$v_3$ to $v_4$ is much smaller than the previous ones.
At $p = 60$ MeV, the variation from ER to
$v_2$ is similar to that from $v_2$ to $v_3$, and the 
variation from $v_3$ to $v_4$ is almost half of $v_2$ and $v_3$.
From around $p \sim 70$ MeV, the 
variation from $v_2$ to $v_3$
exceeds that from ER to $v_2$, and from around $p\sim80$ MeV
the 
variation from $v_3$ to $v_4$ exceeds that from $v_2$ to $v_3$.
Therefore we can say that the effective range expansion up to
$v_4$ converges reasonably up to $p \sim 50$ MeV,
as previously pointed out in Ref.~\cite{ms-plb99}.

\vskip 3mm \noindent
{\bf 6. Discussion and conclusions}

In this work, we calculate the phase shifts of the neutron-proton
scattering for the $^1S_0$ and $^3S_1$ channels at low energies 
employing the di-baryon effective theory with perturbative pions. 
We include the one-pion-exchange diagrams, and expand
their contributions around
the inverse of the tree-level amplitude from the di-baryon fields 
to implement the expansion of the perturbative pions around 
the non-trivial fixed point. 
Loop diagrams are calculated by
using 
DR and the PDS scheme,
and we study our results by changing the value of the 
renormalization scale parameter $\mu$.

We renormalize the scattering length $a$ and the 
effective range $r$ to the physical values.
The scale dependence 
associated with the one-pion-exchange contributions are
encoded in the functions $a_d(\mu)$ and $r_d(\mu)$
which we retain
in the higher order terms of the effective range expansion
as the probe of the optimal higher order corrections.
These functions take part in the physical observable $p \cot \delta_0$
through the term denoted by $F(p)$.
We also calculate the coefficients in the higher-order 
effective range corrections, $v_2$, $v_3$, and $v_4$. 
We obtain new corrections from $r_d(\mu)$ to the terms 
obtained by Cohen and Hansen~\cite{ch1-prc99,ch2-prc99}. 
Those terms do not have unknown constants and are solely determined 
by the terms induced by the one-pion-exchange contributions. 
The $\mu$ dependences of $v_2$, $v_3$, and $v_4$ are plotted and 
we found the sensitivity to $\mu$. 
More sensitive dependence on $\mu$ is observed
in the higher-order term with the units ($fm^3$, $fm^5$, $fm^7$);
$v_4$ is more sensitive than others, $v_3$ is medium, 
and $v_2$ is less sensitive and exhibits a mild $\mu$ dependence. 
By adjusting the $\mu$ value to the known $v_2$ value, however,
we find that we can reasonably reproduce the $v_3$ and $v_4$ 
values obtained from partial wave analysis. 
If $\mu$ value is chosen randomly,
it is hard to say that all the $v_2$, $v_3$ and $v_4$ values are well 
controlled and determined by the terms up to the one-pion-exchange 
contributions.
In addition, this feature can be observed as well
from the $\mu$ dependence of the phase shifts;    
applicable range of the present formalism 
up to the one-pion-exchange corrections is narrow, 
less than $p\sim 50$ MeV with $\mu$ = $m_\pi$, $2m_\pi$ and $3m_\pi$.
However, if we use our best values of $\mu$ for $v_2$, 
then the range of the applicability of the theory 
is considerably widened, up to $p\sim m_\pi$.

Now we briefly discuss the work reported by Soto and Tarrus
employing the di-baryon fields and the perturbative
pions up to NNLO~\cite{st-prc10}.
In that work, the authors included one higher order corrections
than the present work, whereas they employed a slightly different 
counting rule, especially in the leading amplitude, $A_d$ 
in Eq.~(\ref{eq;Ad}), and the different parameter fitting method.
Because they focus on the momentum region, $p\sim m_\pi$, 
contributions from the effective range parameters, $1/a$ and $r$,
in 
$A_d$ are regarded as small corrections
compared to the $ip$ term generated from the nucleon bubble diagram, 
and are expanded
as perturbative terms. Thus scattering amplitudes obtained in this expansion 
scheme do not reproduce the pole structures at very low energies 
in the S-wave nucleon-nucleon scattering, and they have fitted 
parameters by using the phase shifts data at $p=53\sim$ 216~MeV 
($E=3\sim$ 50~MeV). 
In the NNLO calculation, they found similar results to those
obtained in the KSW scheme without the di-baryon fields 
up to NNLO~\cite{fms-npa00}; 
the phase shifts for the $^1S_0$ channel are reproduced well
up to $p\simeq 216$~MeV ($E\simeq 50$~MeV), whereas the calculated
phase shifts for the $^3S_1$ channel deviate from those obtained 
in the accurate potential models at higher momenta and are good 
up to $p\simeq 137$~MeV ($E\simeq 20$~MeV). 
This may imply that the nonperturbative part of the one-pion exchange
contribution in two-pion exchange diagrams is not properly renormalized
in the higher momentum region, $p> 137$~MeV, and the perturbative
treatment in the KSW scheme for the $^3S_1$ channel, even including
the di-baryon field, is stalled~\cite{bbsk-npa02}.
This issue is indeed beyond the scope of the present our work 
including up to the one-pion-exchange contribution,
and we may come back to it in the future.

Though the scale parameter $\mu$ has been conventionally fixed 
as $\mu=m_\pi$, we present that the variation of $\mu$  
provides us a simple and useful test to investigate the range 
of the applicability of the theory.
What we have obtained above may just imply a confirmation of 
a previous result; the slow convergence of the corrections
from the perturbative pions, but we have newly and quantitatively
investigated the range of applicability of the theory as a function 
of the scale parameter $\mu$.

\vskip 3mm \noindent
{\bf Acknowledgements}

We thank Tae-Sun Park for useful comment to our work and discussions.
This work was supported by the Daegu University
Research Grant, 2011.

\end{document}